\documentclass[sigconf]{acmart}
\AtBeginDocument{%
  }

\usepackage{subcaption}
\usepackage{stfloats}

\def\RV{}

\copyrightyear{2026}
\acmYear{2026}
\setcopyright{cc}
\setcctype{by}
\acmConference[CHI EA '26]{Extended Abstracts of the 2026 CHI Conference on Human Factors in Computing Systems}{April 13--17, 2026}{Barcelona, Spain}
\acmBooktitle{Extended Abstracts of the 2026 CHI Conference on Human Factors in Computing Systems (CHI EA '26), April 13--17, 2026, Barcelona, Spain}
\acmDOI{10.1145/3772363.3798826}
\acmISBN{979-8-4007-2281-3/2026/04}


\begin{document}

\title[The Impact of Luminance and Contrast on Reinforcement Learning-based Interaction]{Visual Bias in Simulated Users: The Impact of Luminance and Contrast on Reinforcement Learning-based Interaction}

\author{Hannah Selder}
\orcid{0009-0008-7049-1630} 
\affiliation{%
  \institution{Center for Scalable Data Analytics and Artificial Intelligence (ScaDS.AI) Dresden/Leipzig, Leipzig University}
  \city{Leipzig}
  \country{Germany}
}
\email{hannah.selder@uni-leipzig.de}

\author{Charlotte Beylier}
\orcid{0000-0002-2025-5583} 
\affiliation{%
	\institution{Max Planck Institute for Human Cognitive and Brain Sciences \& Center for Scalable Data Analytics and Artificial Intelligence (ScaDS.AI) Dresden/Leipzig}
	\city{Leipzig}
	\country{Germany}
}
\email{beylier@cbs.mpg.de}

\author{Nico Scherf}
\orcid{0000-0003-4003-9121} 
\affiliation{%
	\institution{Max Planck Institute for Human Cognitive and Brain Sciences \& Center for Scalable Data Analytics and Artificial Intelligence (ScaDS.AI) Dresden/Leipzig}
	\city{Leipzig}
	\country{Germany}
}
\email{nscherf@cbs.mpg.de}

\author{Arthur Fleig}
\orcid{0000-0003-4987-7308} 
\affiliation{%
  \institution{Center for Scalable Data Analytics and Artificial Intelligence (ScaDS.AI) Dresden/Leipzig, Leipzig University}
  \city{Leipzig}
  \country{Germany}
}
\email{arthur.fleig@uni-leipzig.de}

\begin{abstract}
Reinforcement learning (RL) enables simulations of HCI tasks, yet their validity is questionable when performance is driven by visual rendering artifacts distinct from interaction design. We provide the first systematic analysis of how luminance and contrast affect behavior by training 247 \RV{simulated users using RL} on pointing and tracking tasks. We vary the luminance of task-relevant objects, distractors, and background under no distractor, static distractor, and moving distractor conditions, and evaluate task performance and robustness to unseen luminances. Results show luminance becomes critical with static distractors, substantially degrading performance and robustness, whereas motion cues mitigate this issue. Furthermore, robustness depends on preserving relational ordering between luminances rather than matching absolute values. Extreme luminances, especially black, often yield high performance but poor robustness. Overall, seemingly minor luminance changes can strongly shape learned behavior, revealing critical insights into what RL-driven simulated users actually learn.
\end{abstract}



\maketitle

\section{Introduction}
Simulations offer significant potential for Human-Computer Interaction (HCI) by enabling automated, reproducible, and scalable evaluation of interaction techniques and interface designs~\cite{simulations_for_hci_22_murray, simuser_xiang_24}. 
They allow low cost, systematic exploration of large design spaces and can complement empirical evaluations with humans~\cite{simulations_for_hci_22_murray}.
Particularly promising are RL-driven \RV{simulated users}, since they can be trained without human data and, when combined with cognitive models and biomechanical simulations, can ground learned behavior in human constraints~\cite{theory_of_mind_zhu_24, crtypist_24_shi}.

Despite their promise, %
RL-driven \RV{simulated users, or agents,} carry significant unknowns.
While prior work emphasized biomechanical fidelity~\cite{breathing_2022,fischer2021reinforcement,fischer24sim2vr} \RV{and practical guidelines for successful training}~\cite{demystifying2025}, how agents interpret \textit{visual} information remains underexplored. 
Although central to interface design, color in RL environments is often treated as an incidental %
detail rather than a conscious design parameter~\cite{slaoui_robust_19}. 
Recent work shows that CNN-based vision models, used in many RL agents, evaluate interface color in ways aligned with human judgments~\cite{wang2025deep}.
However, %
it is unclear how color influences an agent's behavior, particularly in visually more complex scenes containing distractors, and whether performance remains consistent if the same environment is rendered differently.

If simulated users are sensitive to color changes, performance differences may reflect visual bias rather than interface quality, limiting their validity as "crash test dummies"~\cite{breathing_2022}. %
\textbf{In this work}, we thus systematically analyze this sensitivity \RV{in visuomotor pointing and tracking tasks adapted from the User-in-the-Box framework~\cite{breathing_2022}}. 
To balance computational feasibility with systematic depth -- training a single \RV{simulated user} requires 12 hours of compute -- we operationalize color through luminance and contrast. 
This eliminates confounding chromaticity effects while allowing us to scale our analysis to hundreds of configurations. Guided by this scope, we pose the following research questions:
\begin{itemize}
\item[\textbf{RQ1:}] How do luminance and contrast between task-relevant objects, background, and distractors affect task performance under no-distractor, static-distractor, and moving-distractor conditions in tracking and pointing?
\item[\textbf{RQ2:}] How do these luminance and contrast choices affect the robustness of \RV{the trained simulated users} to changes in visual appearances across these settings?
\end{itemize}
Choosing tracking and pointing as two core HCI tasks, we train \RV{simulated users} across 247 distinct configurations, systematically varying the rendering of targets, backgrounds, and distractors. 
We evaluate both task performance on the trained luminance and robustness across luminances unseen during training. 
We find that performance is stable without distractors but degrades with static distractors unless contrast is high; motion mitigates this sensitivity. 
Robustness %
depends on preservation of relational luminance ordering across training and evaluation rather than absolute values.

We \textbf{contribute to computational modeling} (1) the first systematic analysis of how luminance and contrast affect performance and robustness in RL-driven simulated users, %
\RV{and (2) an openly available dataset\footnote{\url{https://doi.org/10.5281/zenodo.18346733}} of these 247 trained simulated users to facilitate replication and future work on their visual robustness.} %

\begin{figure*}
	\centering
	\begin{subfigure}[t]{0.33\textwidth}
		\centering
		\includegraphics[width=\textwidth]{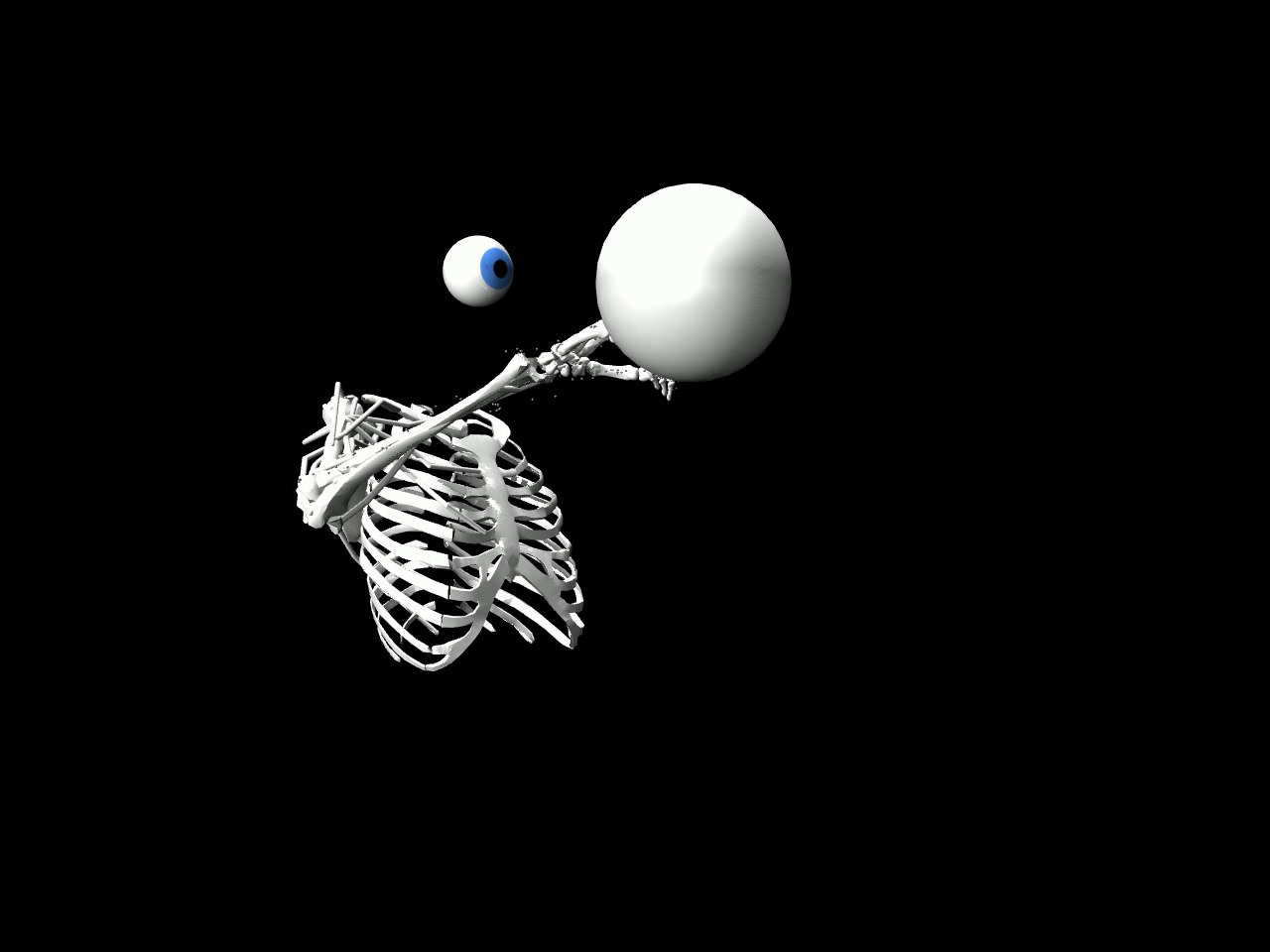}
		\label{sub_fig:pointing_black_back}
	\end{subfigure}
	\begin{subfigure}[t]{0.33\textwidth}
		\centering
		\includegraphics[width=\textwidth]{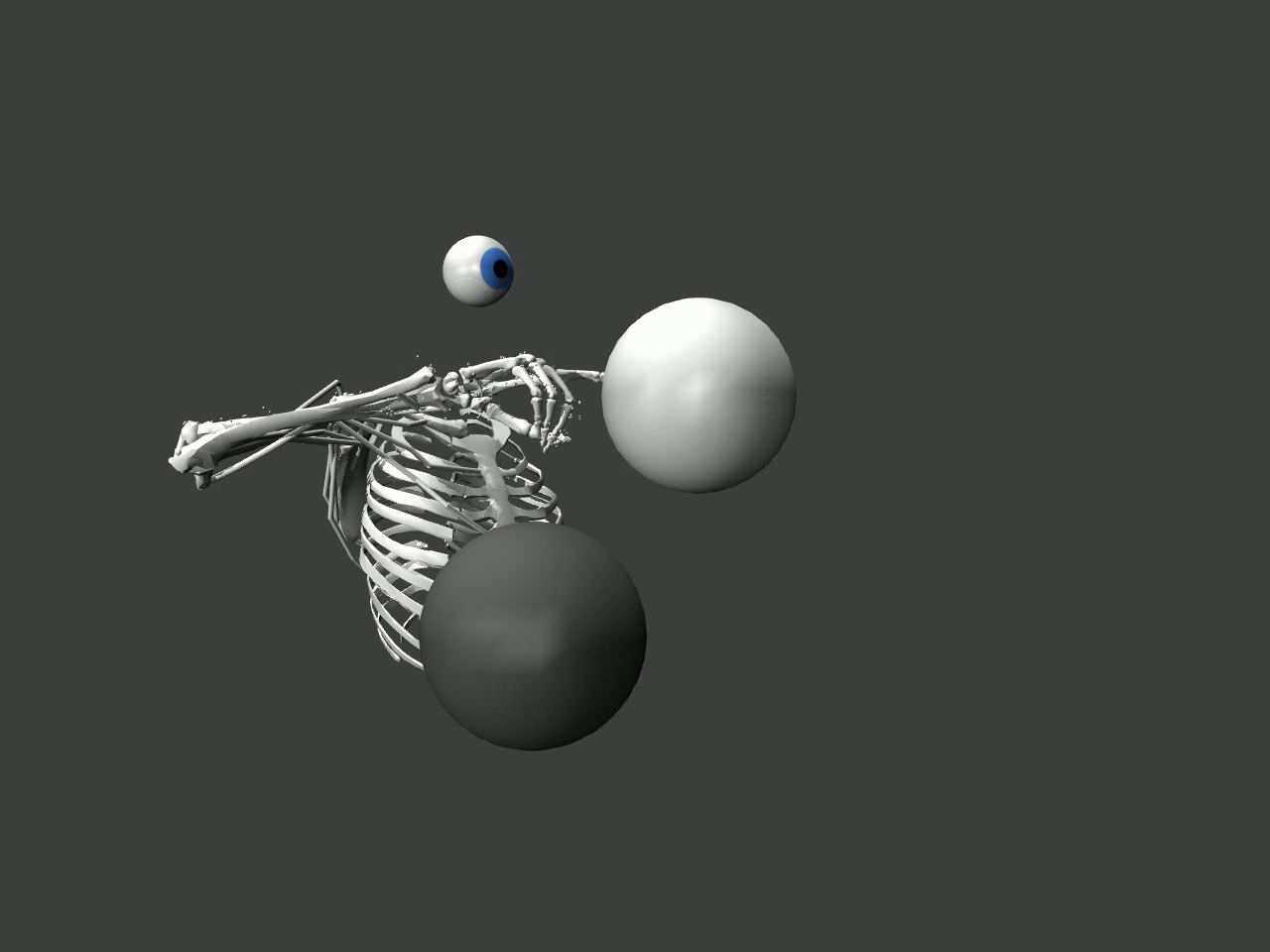}
		\label{sub_fig:pointing_distractor}
	\end{subfigure}
	\begin{subfigure}[t]{0.33\textwidth}
		\centering
		\includegraphics[width=\textwidth]{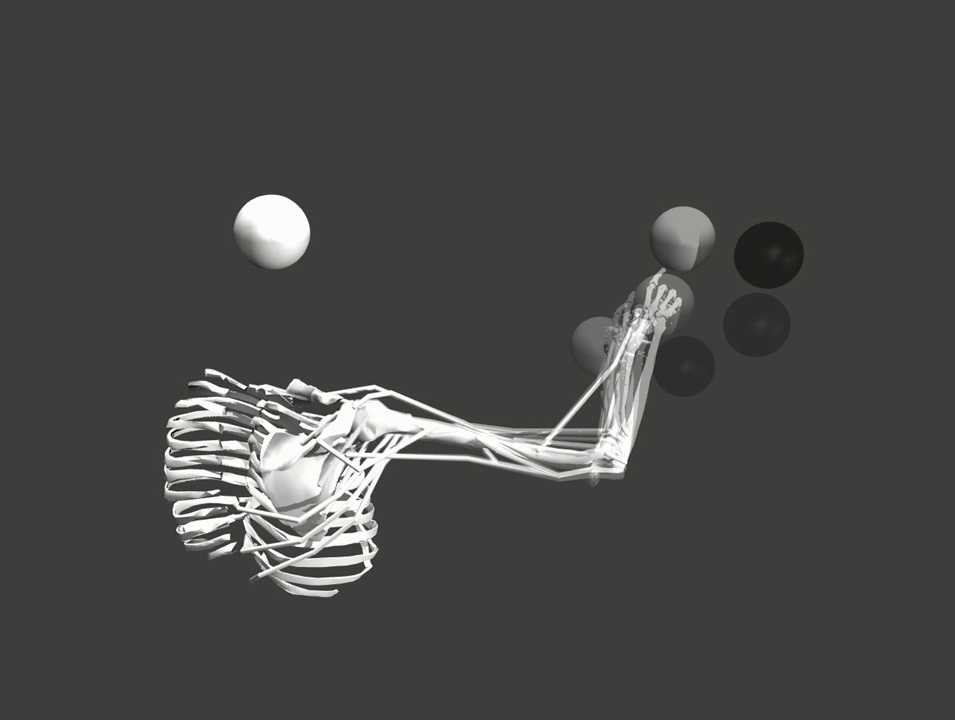}
		\label{sub_fig:tracking_moving_distractor}
	\end{subfigure}\hfill
	\caption{Illustration of the tasks. (a) Pointing task with a black background and white task-relevant objects. (b) Pointing task with a target (bright object) and a distractor (dark object). (c) Tracking task with a moving target and distractor right of it.}
	\Description{The figure shows three side-by-side renderings of the experimental tasks. In all panels, a skeletal human upper body with an eye above and arm extend toward spherical objects in front of it. The left panel depicts a single target sphere positioned in front of the hand, representing a basic pointing scenario without distractors. The middle panel adds a second sphere near the target, illustrating the presence of a distractor during pointing. The right panel shows a tracking scenario in which both target and distractor move; multiple faint, overlaid positions of the spheres indicate motion over time. The arm is oriented toward the target in each case, highlighting the interaction focus of the task.}
	\label{fig:tasks}
\end{figure*}

\section{Related Work}
\textbf{Biomechanical Simulations in HCI. }
Simulations enable automated and scalable evaluation of interaction tasks~\cite{simulations_for_hci_22_murray}. By incorporating physical constraints and motor control, biomechanical simulations model realistic interaction. %
Frameworks such as User-In-The-Box\footnote{\url{https://github.com/User-in-the-Box/user-in-the-box}} combine muscosceletal and perceptual models driven by RL to simulate pointing, tracking, and manipulation tasks~\cite{breathing_2022,fischer24sim2vr}. While significant advances have been made in physical realism and cognitive models~\cite{mind_and_motion_25_fleig}, \textit{visual} properties of simulated environments so far have been treated as incidental, leaving open how visualization choices shape learned interaction behavior.

\textbf{Color and Visual Attention in HCI. }
A core goal of HCI simulation is to model not only human movement but also perception and cognition. Prior work shows that color, through both chromatic and luminance contrast, affects visual attention, discrimination, and task performance~\cite{leiva_understanding_2020, mateescu_attention_2014, shen_15_effect}.
Color-based visualizations can further reduce cognitive load and improve user experience~\cite{He_visualization_25}. In simulation contexts, eye-movement models benefit from color information~\cite{hamel_color_2014}, and CNN-based models exhibit sensitivities aligned with human judgments~\cite{wang2025deep}, raising the question of whether similar alignment emerges in RL-based simulated users. %

\textbf{Visual Robustness and Distractors in RL. }
RL agents in non-biomechanical contexts exhibit systematic sensitivity to visual appearance, including color and background~\cite{Wang_2021_CVPR}. Prior work has explored robustness through domain randomization~\cite{slaoui_robust_19} and by evaluating performance under varying backgrounds and distractors or masking irrelevant features~\cite{grooten_madi_2024, ortiz_dmc-vb_nodate}.
While these approaches primarily aim to reduce the impact of visual variation, color in HCI is an intentional design parameter, particularly in visually rich settings such as VR and AR, where distractors are inherent and color plays a central role in guiding attention and interaction.

\textbf{Research Gap. }
Although color strongly affects human perception and RL agents are sensitive to visual variation, color is rarely studied as a design factor in RL-based interaction simulations. We start by examining how luminance and contrast influences performance and robustness in biomechanical RL agents.

\section{Methodology}
\begin{figure*}
    \begin{minipage}[t]{0.33\textwidth}
        \centering
        \includegraphics[width=\linewidth]{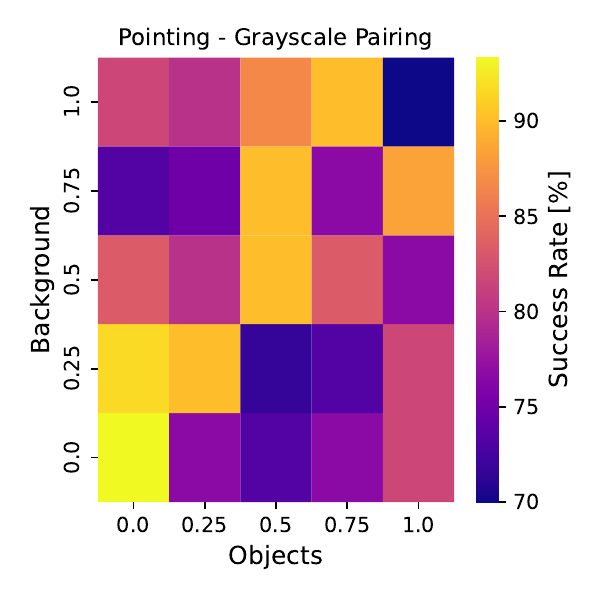}
    \end{minipage}%
    \begin{minipage}[t]{0.33\textwidth}
        \centering
        \includegraphics[width=\linewidth]{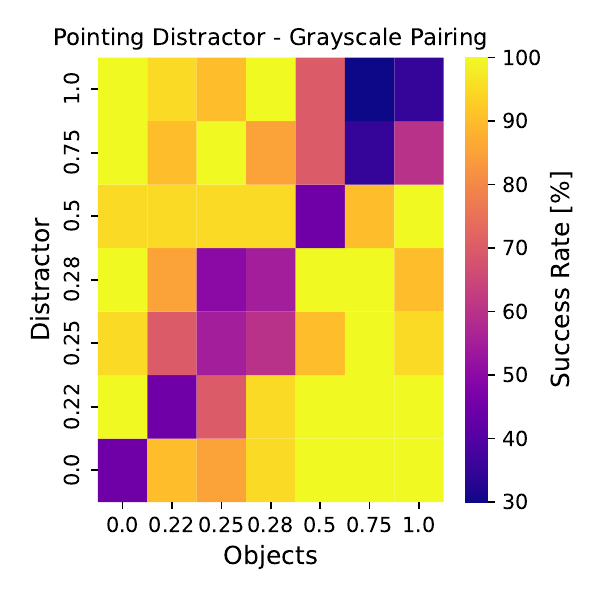}
    \end{minipage}%
    \begin{minipage}[t]{0.33\textwidth}
        \centering
        \includegraphics[width=\linewidth]{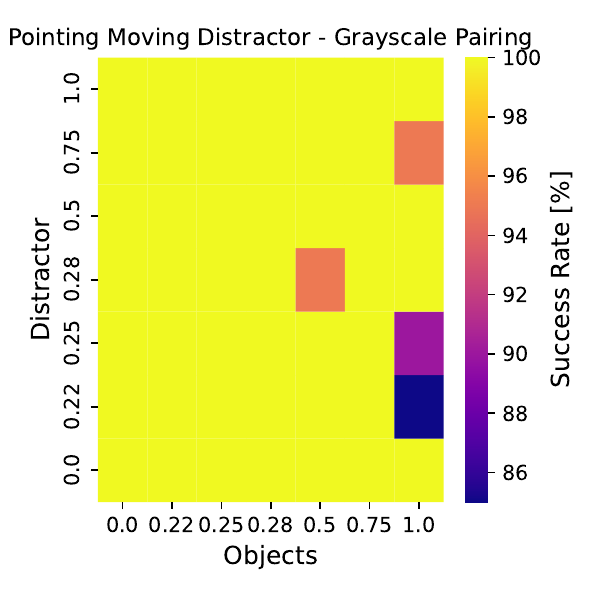}
    \end{minipage}%
    \caption{Pointing performance. Note the different success rate scales. (Left) Performance across background-object luminance combinations. Lower background-object contrast yields better performance. (Middle) Performance across object-static-distractor luminance combinations. (Right) Performance across object-moving-distractor luminance combinations. High object-distractor contrast helps with static distractors, while distractor luminance has virtually no effect when the distractor moves. Background luminance is fixed to 0.25 in the presence of distractors.}
    \Description{The figure contains three heatmaps showing success rates for the pointing task under different luminance pairings, each axis goes from zero to one. The left panel varies background and object luminance with success rates ranging from 70\% to 100\%. Highest success rates cluster along the diagonal, indicating best performance when background and object share similar luminance; the black-on-black condition performs best, while white-on-white shows reduced performance. The middle panel varies object and static distractor luminance with success rates ranging from 30\% to 100\%. Performance decreases as target-distractor contrast decreases. The right panel varies object and moving distractor luminance with success rates ranging from 85\% to 100\%. Here, success rates remain consistently high across combinations.}
    \label{fig:back_object_color_pointing}
\end{figure*}

\textbf{Task and Framework.}  We study the impact of luminance and contrast on simulated user behavior in two fundamental interaction tasks: \textbf{pointing} and \textbf{tracking}, covering interactions with static and moving targets. Building on the \textit{User-In-The-Box} framework~\cite{breathing_2022}, %
we extended the tasks to evaluate the agent under three visual conditions: no distractor, a static distractor \RV{in the same shape as the target}, and a moving distractor that follows a sinusoidal trajectory (Figure~\ref{fig:tasks}). In the tracking task, the distractor mirrors the target’s motion while maintaining a minimum distance of twice the object radius. Distractors are visually present but do not contribute to task success. 
\RV{The visual information the agent receives about its environment}
contains a single frame for static conditions and a short temporal buffer (0.1 s) for moving conditions. 

\textbf{Visual Encoding and Color Conditions.} 
To isolate luminance-based perception effects, visual scenes are rendered in grayscale with zero transparency~\cite{morrone_color_2002}. \RV{Frames} consist of $120 \times 80$ grayscale images, normalized to the range $[0,1]$, where 0 denotes black and 1 denotes white. %
We group scene elements into three semantic categories: (a) background, (b) task-relevant objects (targets and the biomechanical model), and (c) distractors. Each category is assigned luminances sampled from the full intensity range, enabling systematic manipulation of contrast. 
Only in conditions with distractors, the background luminance is fixed at $0.25$, \RV{in which case we additionally consider luminances near $0.25$.}

\textbf{Evaluation Metrics.} We evaluate both task performance and robustness to changes in luminance.
For pointing, performance is measured by success rate, defined as the proportion of trials in which the agent maintains its fingertip within the (penetrable) target for a predefined dwell time. 
\RV{We evaluate each agent on a total of 20 trials, where each trial consists of pointing to a target randomly spawned within a $30\times 30$cm square 55cm in front of the agent. 
At the beginning of each trial, the agent's arm is hanging down. 
The next trial begins after successful pointing or after a maximum of 4~seconds.}
For tracking, performance is quantified by the average distance between the fingertip and the target \RV{center} over a duration of 10 seconds\RV{, with the agent's arm hanging down at the start}. 
Robustness is evaluated using the same metrics under combinations of task-relevant objects, background, and distractor luminances not seen during training. %

\textbf{RL setup.} Simulated users are trained using the Proximal Policy Optimization (PPO) algorithm~\cite{ppo_schulman}. %
For the biomechanical model, we use \citeauthor{breathing_2022}'s~\cite{breathing_2022} \textit{MoblArmsIndex}, which is the reduced MuJoCo implementation of the state-of-the-art \emph{MoblArms} upper-extremity biomechanical model~\cite{saul2015benchmarking}. \RV{This model represents the upper upper extremity through joints and muscles, translating muscle activations into physically realistic movements.}
For both tasks, we use the respective default reward function from User-in-the-Box~\cite{breathing_2022}, which combines the distance between the end-effector and the target, a penalty on neural effort~\cite{neural_berret_11}, and for pointing a bonus for keeping the fingertip inside the target for the required dwell time. Agents are trained for 25 million steps for the tracking task and 14 million steps for the pointing task, which we found sufficient for convergence. To ensure random effects do not confound our analysis, we repeated first experiments involving background and task-relevant object colors for the pointing task with three different neural network initializations (seeds). Finding minimal deviations, we refrained from running all configurations with three seeds. 

\section{Results}
We report results for agents trained under various luminance combinations, analyzing both task performance and robustness to luminance changes. We illustrate a representative subset of results here and provide further figures in the appendix. %
Throughout, for brevity we refer to task-relevant objects simply as \textit{objects}.

\subsection{Performance}

\textbf{Pointing. }
For background-object combinations (Figure~\ref{fig:back_object_color_pointing} left), the highest success rates cluster along the diagonal, indicating best performance when background and object share similar luminance.
\RV{While the black‑on‑black configuration (0.0, 0.0) performs best, white-on-white (1.0, 1.0) shows clear performance drops, down to $70\%$ success rate.} %
With a \textbf{static distractor} (Figure~\ref{fig:back_object_color_pointing} middle), performance declines as target-distractor contrast decreases. Notably, luminance is the sole available cue to distinguishing the target from the distractor. In contrast, with a \textbf{moving distractor} (Figure~\ref{fig:back_object_color_pointing} right), the additional motion cue makes object luminance largely irrelevant to performance.

\begin{figure*}[!ht]
    \begin{minipage}[t]{0.33\textwidth}
        \centering
        \includegraphics[width=\linewidth]{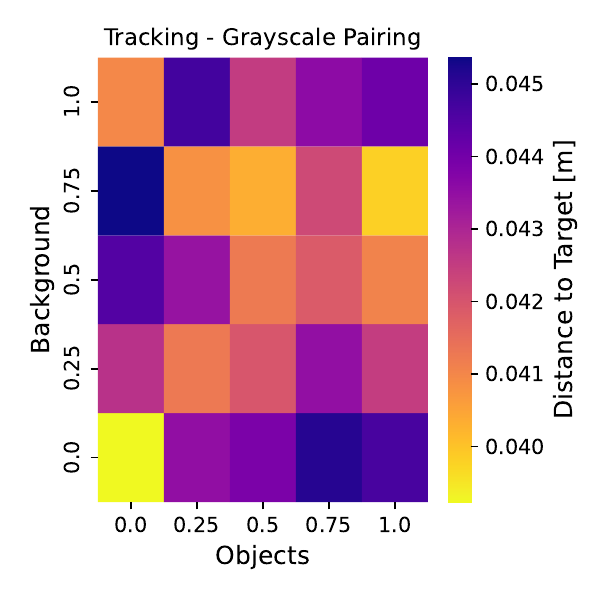}
    \end{minipage}%
    \begin{minipage}[t]{0.33\textwidth}
        \centering
        \includegraphics[width=\linewidth]{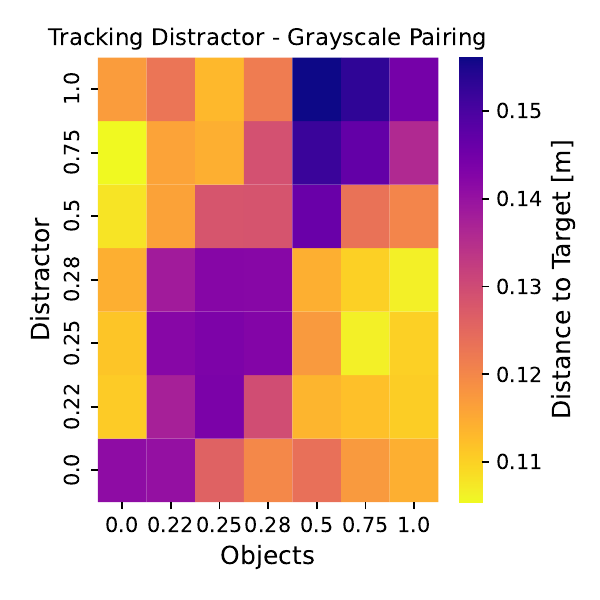}
    \end{minipage}
    \begin{minipage}[t]{0.33\textwidth}
        \centering
        \includegraphics[width=\linewidth]{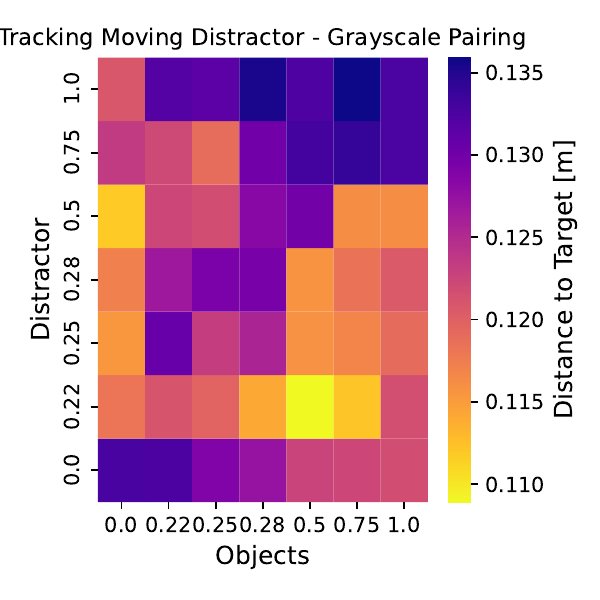}
    \end{minipage}%
    \caption{Tracking performance (lower distance = better). Note the different distance-to-target scales. (Left) %
    Black background and object yield the best performance, with little overall deviation. (Middle) %
    High contrast between objects and static distractors improves performance. (Right) In performance across object-moving-distractor luminance combinations, luminance has little effect.
    We recall that background luminance is fixed to 0.25 in the presence of distractors.}
    
    \Description{Three heatmaps show distance to target in meters for the tracking task under different luminance pairings. In all panels, both axes range from zero to one. Lower values indicate better performance. The left panel varies background and object luminance, with distances ranging from 0.039 m to 0.046 m. The black-on-black condition yields the lowest distance, while white-on-white shows reduced performance; no consistent pattern is visible otherwise. The middle panel varies object and static distractor luminance, with distances ranging from 0.1 m to 0.16 m. Lower distances occur when contrast between object and static distractor is high. The right panel varies object and moving distractor luminance, with distances ranging from 0.11 m to 0.1355 m. Distances tend to be lower at higher object–distractor contrast, though differences are smaller than in the static condition.}
    \label{fig:back_object_color_tracking}
\end{figure*}
\textbf{Tracking. }
For background-object combinations (Figure~\ref{fig:back_object_color_tracking} left), black‑on‑black (0.0, 0.0) again yields the best performance, with no consistent trend otherwise. With a \textbf{static distractor} (Figure~\ref{fig:back_object_color_tracking} middle), performance is lower (different scale). Similar to pointing, performance degrades as target-distractor contrast decreases, but tracking requires higher contrast than pointing for reliable performance. As in pointing, a \textbf{moving distractor} (Figure~\ref{fig:back_object_color_tracking} right) renders object luminance less important\RV{, while performance tends to be better at higher object-distractor contrast.}

\subsection{Robustness}
\begin{figure*}[t]
    \begin{minipage}[t]{0.49\textwidth}
        \centering
        \includegraphics[width=\linewidth]{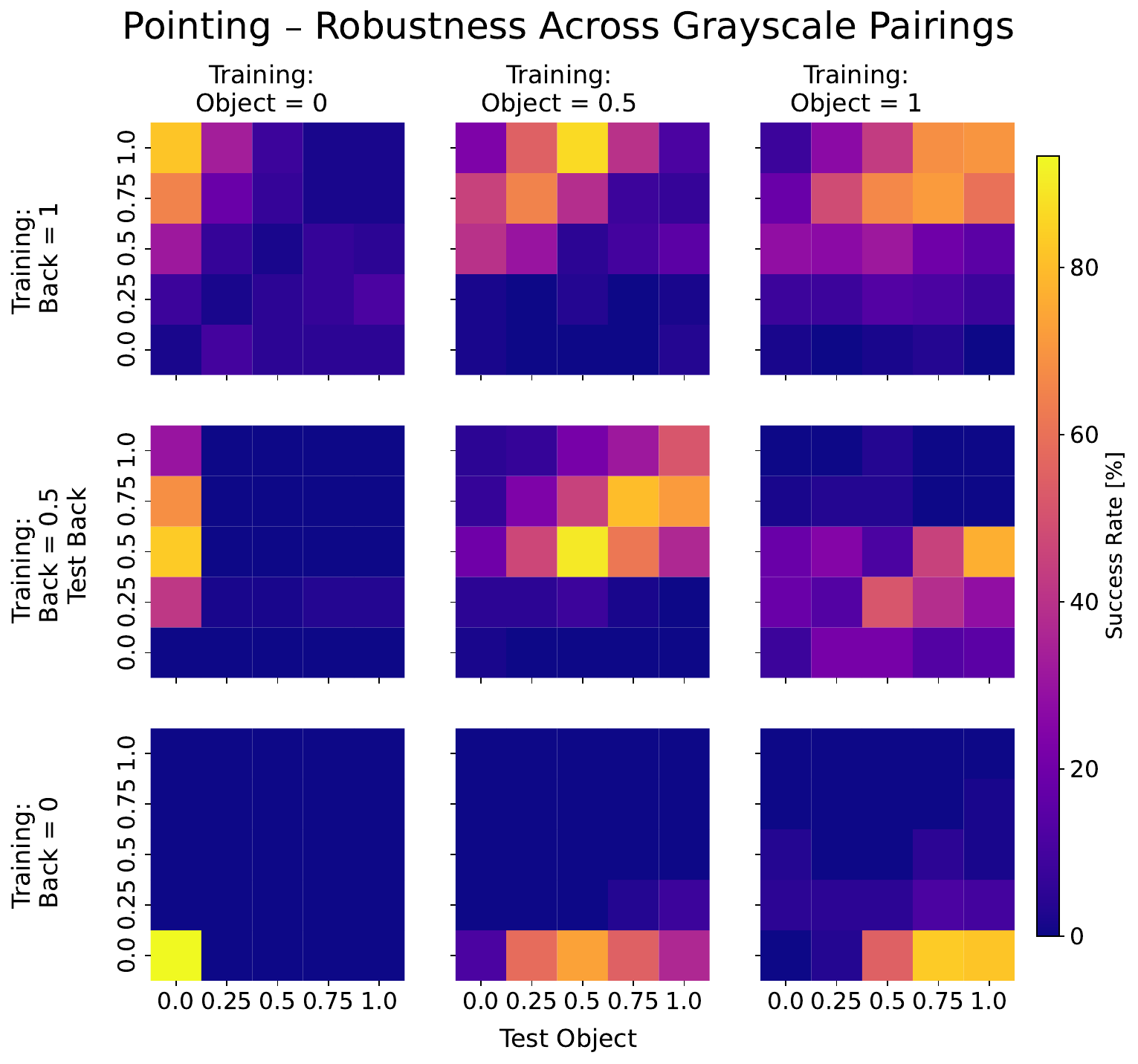}
    \end{minipage}%
    \hfill
    \begin{minipage}[t]{0.49\textwidth}
        \centering
        \includegraphics[width=\linewidth]{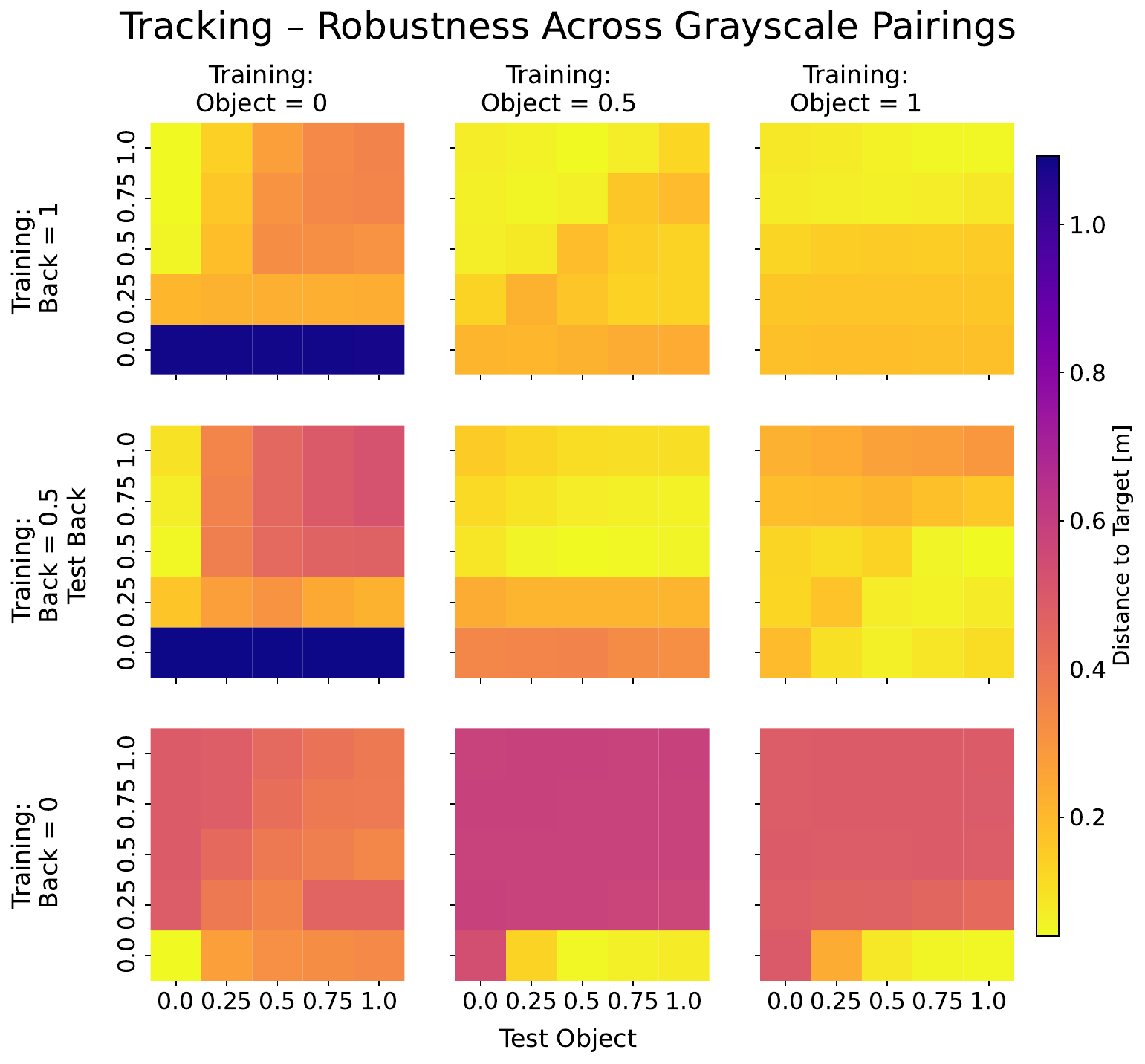}
    \end{minipage}%
    \caption{Robustness across background-object luminance combinations for (left) pointing and (right) tracking. Agents trained on black-on-black achieve high training performance but exhibit poor robustness. Robustness improves when the relational luminance ordering learned during training is preserved: for heatmaps in the upper left, in each individual heatmap the best performance is also in the upper left; analogously for heatmaps in the lower right.} 
    \Description{The two figures present heatmaps showing how agents trained on specific background-object luminance pairings generalize to other luminance combinations. The left figure shows pointing, the right one tracking and each of them consist of 9 heatmaps representing the performance of one agent trained on a specific background-object luminance combination evaluated across different test luminance combinations. Rows indicate the background luminance, and columns show object luminance.
    For pointing, success rates range from 0 to 100\% and agents trained on black background and black object achieve high training performance but generalize poorly, with performance dropping sharply under luminance changes. Robustness is highest when the relational ordering between object and background luminance learned during training is preserved. For tracking, distance to target range from 0.1 m to 1.1 m, and robustness is generally higher, but similar patterns appear: agents generalize best when evaluation luminance combinations maintain the same relative ordering as during training.}
    \label{fig:robust_back_object_color_tracking}
\end{figure*}

\textbf{Pointing. }
Figure~\ref{fig:robust_back_object_color_tracking} (left) shows robustness results, where each panel contains heatmaps of a single agent evaluated across luminance combinations.
The agent achieving the highest training performance (black object on black background; Figure~\ref{fig:back_object_color_pointing} left) is the least robust, as even minor luminance changes collapse pointing success. Robustness is highest when the relational ordering learned during training, $value_{object} < value_{background}$, is preserved, while deviations from black (zero) for either element sharply degrade performance.
With a \textbf{static distractor} (Figure~\ref{fig:pointing_distractor_robustness_heatmap_reduced} in appendix), robustness is maintained when object luminances are close to the background luminance (0.25) and evaluation luminances remain near training values. 
Agents trained with black objects again show the weakest robustness, with performance dropping sharply under larger luminance shifts (e.g., from 0.28 to 0.5). Preserving the ordering $value_{object} < value_{distractor}$ again improves robustness.
With a \textbf{moving distractor} (Figure~\ref{fig:pointing_moving_distractor_robustness_heatmap_reduced} in appendix), robustness patterns largely mirror the static case\RV{, albeit with generally higher success rates. Performance remains stable across distractor luminances, but drops when the object luminance differs too much from the one the agent was trained on.} %

\textbf{Tracking. }
\RV{Compared to pointing,} tracking (Figure~\ref{fig:robust_back_object_color_tracking} right) is generally more robust, though similar failure modes remain. The best-performing configuration (black on black; Figure~\ref{fig:back_object_color_tracking} left) again exhibits the weakest robustness. 
Furthermore, agents trained with black objects or background tend to be less robust. 
With a \textbf{static distractor} (Figure \ref{fig:tracking_distractor_robustness_heatmap_reduced} in appendix), \RV{similar to the moving distractor case in pointing, performance tends to remain stable across distractor luminances and tends to drop when the object luminance deviates too much from the one during training.} 
Agents trained with black objects are particularly brittle. Again, %
preserving the ordering of these luminances improves robustness. 
With a \textbf{moving distractor} (Figure \ref{fig:tracking_moving_distractor_robustness_heatmap_reduced} in appendix), robustness follows the same qualitative pattern as with the static distractor. Black objects generalize poorly, and larger object luminance shifts from the trained luminance lead to rapid performance degradation.

\section{Discussion}

\textbf{Effect of luminance and contrast on performance (RQ1).}
Luminance systematically affects the agent's performance.
Without distractors, agents learn pointing (70\%-100\% success rate) and tracking across all luminance combinations. 
Introducing distractors can lower performance particularly in tracking, and increases perceptual demands:
with static distractors, higher contrast between objects and distractors substantially improves performance, especially in pointing, where luminance is the only cue. 
When distractors are moving, motion provides an additional segmentation signal, reducing reliance on luminance and largely mitigating its impact on performance.
Most notably, for both pointing and tracking, a simulated user trained on a black background is effectively a different agent than one trained on a gray background. 
For HCI practitioners, this creates a risk of evaluation artifacts: a simulated user might erroneously predict higher usability for an interface rendered on a pure black background compared to one with a textured or gray background, purely due to the agent's optimization bias rather than the interaction design itself.

\textbf{Effect of luminance and contrast on robustness (RQ2).}
Robustness depends largely on preserving the \textit{relational ordering} of luminances, not on matching absolute values. Training on extreme luminances, particularly black objects, causes overfitting and sharp drops under minor changes.
However, motion cues mitigate this fragility: when targets or distractors move, agents rely less on contrast. This suggests that dynamic tasks inherently produce more valid simulated users than static tasks, as they force the agent to learn temporal interactions that generalize better than static pattern matching.
A \textit{potential} explanation for why black backgrounds yield high performance but poor robustness is that agents might learn a trivial shortcut: treating any non-zero pixel as "object" and zero as "background". This binary classification is computationally simpler than learning edge detection. However, this shortcut immediately fails when the background becomes non-zero (e.g., gray), as it changes the learned object-background labeling. This mechanism would explain both the accelerated training performance and the catastrophic failure under even minor luminance shifts.

\textbf{Implications and Recommendations. }
Our results suggest that visual rendering deserves more careful attention than it typically receives in RL-based simulation.
Agents may overfit to rendering artifacts (e.g., whether the background is black or gray), and behave in non-obvious ways, such as profiting from preserving the relational contrast ordering from the training environment. 
They can be sensitive to visual rendering in ways that threaten their validity as evaluation proxies, e.g., when simulated users are evaluated across visual themes such as dark vs.\ light mode. %
Even when not caring about robustness and specific colors or luminances, different visual environments lead to different performance outcomes. %
This provides yet another diagnostic axis for researchers working with simulated users to consider, as poor performance might not only result from poorly chosen reward functions or the (lack of) training curricula, but simply because of the chosen rendering. 
For visuomotor tasks in particular, our results reinforce the need to better understand \textit{what} an agent is learning during training~\cite{beylier2025attentiontrajectoriesdiagnosticaxis}.

As a general precaution in interface design, we recommend treating any simulated user trained on a single fixed visual configuration with skepticism, and validating learned agents under at least a small set of luminance variations before drawing conclusions about interface quality. 
To improve robustness, we recommend avoiding training agents solely on pure black or white backgrounds and objects, and instead use mid-range luminance values. 
While one could consider various visual configurations already during training, this might prolong training time, and requires further research. 
A potential silver lining is that motion effectively helps as an additional cue, for both moving targets and distractors, and improves robustness. 
This makes simulated users more usable for real-world AR interfaces with, e.g., moving elements in the background. 
More fundamentally, however, our results suggest that to serve as valid evaluation tools, deeper inquiry is required into how to model perception in a way that enables more human-like evaluations of interfaces.

\textbf{Limitations and Future Work. }
Our study focused on grayscale luminance to isolate contrast effects in a systematic way while balancing computational cost; future work should investigate whether chromaticity introduces similar biases or offers additional cues. 
Similarly, while we considered static and moving distractors, their size, shape, and movement patterns were fixed. 
Further research is required on how these factors influence results. 
More fundamentally, while User-in-the-Box is a state-of-the-art simulation framework for visuomotor HCI interaction tasks, in this first analysis we did not vary its visual encodings; a different vision model might lead to different outcomes. 
Overall, a particularly important direction for future work is to examine how the observed color sensitivities relate to human perceptual behavior, to clarify the validity of simulated users as crash test dummies.

\section{Conclusion}
We studied how luminance and contrast affect performance and robustness in RL-based interaction simulations. Across
247 simulated users trained on pointing and tracking tasks, we showed that luminance systematically affects both task success
and generalization. Our findings have direct implications for the validity of computational interaction research. If
simulated users overfit to rendering artifacts, they cannot reliably distinguish between good and poor interface designs,
undermining their utility as evaluation tools. As RL-driven simulations become more widely adopted in HCI, ensuring
they evaluate interaction quality rather than rendering artifacts is essential for their validity.

\begin{acks}
    The authors acknowledge the financial support by the Federal Ministry of Research, Technology and Space of Germany and by Sächsische Staatsministerium für Wissenschaft, Kultur und Tourismus in the programme Center of Excellence for AI-research „Center for Scalable Data Analytics and Artificial Intelligence Dresden/Leipzig“, project identification number: ScaDS.AI.
    The authors gratefully acknowledge the computing time made available to them on the high-performance computer at the NHR Center of TU Dresden. This center is jointly supported by the Federal Ministry of Research, Technology and Space of Germany and the state governments participating in the NHR (www.nhr-verein.de/unsere-partner).
\end{acks}

\balance
\bibliographystyle{ACM-Reference-Format}
\bibliography{sample-base}


\begin{thebibliography}{24}

%%% the Web address.

\ifx \showCODEN    \undefined \def \showCODEN     #1{\unskip}     \fi
\ifx \showISBNx    \undefined \def \showISBNx     #1{\unskip}     \fi
\ifx \showISBNxiii \undefined \def \showISBNxiii  #1{\unskip}     \fi
\ifx \showISSN     \undefined \def \showISSN      #1{\unskip}     \fi
\ifx \showLCCN     \undefined \def \showLCCN      #1{\unskip}     \fi
\ifx \shownote     \undefined \def \shownote      #1{#1}          \fi
\ifx \showarticletitle \undefined \def \showarticletitle #1{#1}   \fi
\ifx \showURL      \undefined \def \showURL       {\relax}        \fi
% invisible to TeX
\providecommand\bibfield[2]{#2}
\providecommand\bibinfo[2]{#2}
\providecommand\natexlab[1]{#1}
\providecommand\showeprint[2][]{arXiv:#2}

\bibitem[Berret et~al\mbox{.}(2011)]%
        {neural_berret_11}
\bibfield{author}{\bibinfo{person}{Bastien Berret}, \bibinfo{person}{Enrico
  Chiovetto}, \bibinfo{person}{Francesco Nori}, {and} \bibinfo{person}{Thierry
  Pozzo}.} \bibinfo{year}{2011}\natexlab{}.
\newblock \showarticletitle{Evidence for Composite Cost Functions in Arm
  Movement Planning: An Inverse Optimal Control Approach}.
\newblock \bibinfo{journal}{\emph{PLoS computational biology}}
  \bibinfo{volume}{7} (\bibinfo{date}{10} \bibinfo{year}{2011}),
  \bibinfo{pages}{e1002183}.
\newblock
\href{https://doi.org/10.1371/journal.pcbi.1002183}{doi:\nolinkurl{10.1371/journal.pcbi.1002183}}


\bibitem[Beylier et~al\mbox{.}(2025)]%
        {beylier2025attentiontrajectoriesdiagnosticaxis}
\bibfield{author}{\bibinfo{person}{Charlotte Beylier}, \bibinfo{person}{Hannah
  Selder}, \bibinfo{person}{Arthur Fleig}, \bibinfo{person}{Simon~M. Hofmann},
  {and} \bibinfo{person}{Nico Scherf}.} \bibinfo{year}{2025}\natexlab{}.
\newblock \bibinfo{title}{Attention Trajectories as a Diagnostic Axis for Deep
  Reinforcement Learning}.
\newblock
\showeprint[arxiv]{2511.20591}~[cs.LG]
\urldef\tempurl%
\url{https://arxiv.org/abs/2511.20591}
\showURL{%
\tempurl}


\bibitem[Fischer et~al\mbox{.}(2021)]%
        {fischer2021reinforcement}
\bibfield{author}{\bibinfo{person}{Florian Fischer}, \bibinfo{person}{Miroslav
  Bachinski}, \bibinfo{person}{Markus Klar}, \bibinfo{person}{Arthur Fleig},
  {and} \bibinfo{person}{J{\"o}rg M{\"u}ller}.}
  \bibinfo{year}{2021}\natexlab{}.
\newblock \showarticletitle{Reinforcement learning control of a biomechanical
  model of the upper extremity}.
\newblock \bibinfo{journal}{\emph{Scientific Reports}} \bibinfo{volume}{11},
  \bibinfo{number}{1} (\bibinfo{year}{2021}), \bibinfo{pages}{14445}.
\newblock


\bibitem[Fischer et~al\mbox{.}(2024)]%
        {fischer24sim2vr}
\bibfield{author}{\bibinfo{person}{Florian Fischer}, \bibinfo{person}{Aleksi
  Ikkala}, \bibinfo{person}{Markus Klar}, \bibinfo{person}{Arthur Fleig},
  \bibinfo{person}{Miroslav Bachinski}, \bibinfo{person}{Roderick
  Murray-Smith}, \bibinfo{person}{Perttu H\"{a}m\"{a}l\"{a}inen},
  \bibinfo{person}{Antti Oulasvirta}, {and} \bibinfo{person}{J\"{o}rg
  M\"{u}ller}.} \bibinfo{year}{2024}\natexlab{}.
\newblock \showarticletitle{SIM2VR: Towards Automated Biomechanical Testing in
  VR}. In \bibinfo{booktitle}{\emph{Proceedings of the 37th Annual ACM
  Symposium on User Interface Software and Technology}} (Pittsburgh, PA, USA)
  \emph{(\bibinfo{series}{UIST '24})}. \bibinfo{publisher}{Association for
  Computing Machinery}, \bibinfo{address}{New York, NY, USA}, Article
  \bibinfo{articleno}{79}, \bibinfo{numpages}{15}~pages.
\newblock
\showISBNx{9798400706288}
\href{https://doi.org/10.1145/3654777.3676452}{doi:\nolinkurl{10.1145/3654777.3676452}}


\bibitem[Fleig et~al\mbox{.}(2025)]%
        {mind_and_motion_25_fleig}
\bibfield{author}{\bibinfo{person}{Arthur Fleig}, \bibinfo{person}{Florian
  Fischer}, \bibinfo{person}{Markus Klar}, \bibinfo{person}{Patrick Ebel},
  \bibinfo{person}{Miroslav Bachinski}, \bibinfo{person}{Per~Ola Kristensson},
  \bibinfo{person}{Roderick Murray-Smith}, {and} \bibinfo{person}{Antti
  Oulasvirta}.} \bibinfo{year}{2025}\natexlab{}.
\newblock \showarticletitle{Mind \& Motion: Opportunities and Applications of
  Integrating Biomechanics and Cognitive Models in HCI}. In
  \bibinfo{booktitle}{\emph{Adjunct Proceedings of the 38th Annual ACM
  Symposium on User Interface Software and Technology}}
  \emph{(\bibinfo{series}{UIST Adjunct '25})}. \bibinfo{publisher}{Association
  for Computing Machinery}, \bibinfo{address}{New York, NY, USA}, Article
  \bibinfo{articleno}{14}, \bibinfo{numpages}{4}~pages.
\newblock
\showISBNx{9798400720369}
\href{https://doi.org/10.1145/3746058.3758473}{doi:\nolinkurl{10.1145/3746058.3758473}}


\bibitem[Grooten et~al\mbox{.}(2024)]%
        {grooten_madi_2024}
\bibfield{author}{\bibinfo{person}{Bram Grooten}, \bibinfo{person}{Tristan
  Tomilin}, \bibinfo{person}{Gautham Vasan}, \bibinfo{person}{Matthew~E.
  Taylor}, \bibinfo{person}{A.~Rupam Mahmood}, \bibinfo{person}{Meng Fang},
  \bibinfo{person}{Mykola Pechenizkiy}, {and}
  \bibinfo{person}{Decebal~Constantin Mocanu}.}
  \bibinfo{year}{2024}\natexlab{}.
\newblock \showarticletitle{MaDi: Learning to Mask Distractions for
  Generalization in Visual Deep Reinforcement Learning}. In
  \bibinfo{booktitle}{\emph{Proceedings of the 23rd International Conference on
  Autonomous Agents and Multiagent Systems}} (Auckland, New Zealand)
  \emph{(\bibinfo{series}{AAMAS '24})}. \bibinfo{publisher}{International
  Foundation for Autonomous Agents and Multiagent Systems},
  \bibinfo{address}{Richland, SC}, \bibinfo{pages}{733–742}.
\newblock
\showISBNx{9798400704864}


\bibitem[Hamel et~al\mbox{.}(2014)]%
        {hamel_color_2014}
\bibfield{author}{\bibinfo{person}{Shahrbanoo Hamel}, \bibinfo{person}{Nathalie
  Guyader}, \bibinfo{person}{Denis Pellerin}, {and} \bibinfo{person}{Dominique
  Houzet}.} \bibinfo{year}{2014}\natexlab{}.
\newblock \showarticletitle{{Color Information in a Model of Saliency}}. In
  \bibinfo{booktitle}{\emph{{EUSIPCO 2014 Conference Proceedings}}}.
  \bibinfo{publisher}{Institute of Electrical and Electronics Engineers
  (IEEE)}, \bibinfo{address}{Lisbonne, Portugal}, \bibinfo{pages}{1--5}.
\newblock
\urldef\tempurl%
\url{https://hal.science/hal-01068276}
\showURL{%
\tempurl}


\bibitem[He et~al\mbox{.}(2025)]%
        {He_visualization_25}
\bibfield{author}{\bibinfo{person}{Xiaotong He}, \bibinfo{person}{Yuqian Yang},
  \bibinfo{person}{Ruojuan Li}, {and} \bibinfo{person}{Xiaodong Gong}.}
  \bibinfo{year}{2025}\natexlab{}.
\newblock \showarticletitle{Visualization Design and Color Preference in
  Medical Application Interfaces for Older Adults: The Impact on Cognitive Load
  and User Experience}.
\newblock \bibinfo{journal}{\emph{International Journal of Human–Computer
  Interaction}} \bibinfo{volume}{0}, \bibinfo{number}{0}
  (\bibinfo{year}{2025}), \bibinfo{pages}{1--21}.
\newblock
\href{https://doi.org/10.1080/10447318.2025.2562959}{doi:\nolinkurl{10.1080/10447318.2025.2562959}}


\bibitem[Ikkala et~al\mbox{.}(2022)]%
        {breathing_2022}
\bibfield{author}{\bibinfo{person}{Aleksi Ikkala}, \bibinfo{person}{Florian
  Fischer}, \bibinfo{person}{Markus Klar}, \bibinfo{person}{Miroslav
  Bachinski}, \bibinfo{person}{Arthur Fleig}, \bibinfo{person}{Andrew Howes},
  \bibinfo{person}{Perttu H\"{a}m\"{a}l\"{a}inen}, \bibinfo{person}{J\"{o}rg
  M\"{u}ller}, \bibinfo{person}{Roderick Murray-Smith}, {and}
  \bibinfo{person}{Antti Oulasvirta}.} \bibinfo{year}{2022}\natexlab{}.
\newblock \showarticletitle{Breathing Life Into Biomechanical User Models}. In
  \bibinfo{booktitle}{\emph{Proceedings of the 35th Annual ACM Symposium on
  User Interface Software and Technology}} (Bend, OR, USA)
  \emph{(\bibinfo{series}{UIST '22})}. \bibinfo{publisher}{Association for
  Computing Machinery}, \bibinfo{address}{New York, NY, USA}, Article
  \bibinfo{articleno}{90}, \bibinfo{numpages}{14}~pages.
\newblock
\showISBNx{9781450393201}
\href{https://doi.org/10.1145/3526113.3545689}{doi:\nolinkurl{10.1145/3526113.3545689}}


\bibitem[Leiva et~al\mbox{.}(2020)]%
        {leiva_understanding_2020}
\bibfield{author}{\bibinfo{person}{Luis~A. Leiva}, \bibinfo{person}{Yunfei
  Xue}, \bibinfo{person}{Avya Bansal}, \bibinfo{person}{Hamed~R. Tavakoli},
  \bibinfo{person}{Tuðçe Köroðlu}, \bibinfo{person}{Jingzhou Du},
  \bibinfo{person}{Niraj~R. Dayama}, {and} \bibinfo{person}{Antti Oulasvirta}.}
  \bibinfo{year}{2020}\natexlab{}.
\newblock \showarticletitle{Understanding {Visual} {Saliency} in {Mobile}
  {User} {Interfaces}}. In \bibinfo{booktitle}{\emph{22nd {International}
  {Conference} on {Human}-{Computer} {Interaction} with {Mobile} {Devices} and
  {Services}}}. \bibinfo{publisher}{ACM}, \bibinfo{address}{Oldenburg Germany},
  \bibinfo{pages}{1--12}.
\newblock
\showISBNx{978-1-4503-7516-0}
\href{https://doi.org/10.1145/3379503.3403557}{doi:\nolinkurl{10.1145/3379503.3403557}}


\bibitem[Mateescu and Bajić(2014)]%
        {mateescu_attention_2014}
\bibfield{author}{\bibinfo{person}{Victor~A. Mateescu} {and}
  \bibinfo{person}{Ivan~V. Bajić}.} \bibinfo{year}{2014}\natexlab{}.
\newblock \showarticletitle{Attention {Retargeting} by {Color} {Manipulation}
  in {Images}}. In \bibinfo{booktitle}{\emph{Proceedings of the 1st
  {International} {Workshop} on {Perception} {Inspired} {Video} {Processing}}}.
  \bibinfo{publisher}{ACM}, \bibinfo{address}{Orlando Florida USA},
  \bibinfo{pages}{15--20}.
\newblock
\showISBNx{978-1-4503-3125-8}
\href{https://doi.org/10.1145/2662996.2663009}{doi:\nolinkurl{10.1145/2662996.2663009}}


\bibitem[Morrone et~al\mbox{.}(2002)]%
        {morrone_color_2002}
\bibfield{author}{\bibinfo{person}{Maria~Concetta Morrone},
  \bibinfo{person}{Valentina Denti}, {and} \bibinfo{person}{Donatella
  Spinelli}.} \bibinfo{year}{2002}\natexlab{}.
\newblock \showarticletitle{Color and {Luminance} {Contrasts} {Attract}
  {Independent} {Attention}}.
\newblock \bibinfo{journal}{\emph{Current Biology}} \bibinfo{volume}{12},
  \bibinfo{number}{13} (\bibinfo{date}{July} \bibinfo{year}{2002}),
  \bibinfo{pages}{1134--1137}.
\newblock
\showISSN{09609822}
\href{https://doi.org/10.1016/S0960-9822(02)00921-1}{doi:\nolinkurl{10.1016/S0960-9822(02)00921-1}}


\bibitem[Murray-Smith et~al\mbox{.}(2022)]%
        {simulations_for_hci_22_murray}
\bibfield{author}{\bibinfo{person}{Roderick Murray-Smith},
  \bibinfo{person}{Antti Oulasvirta}, \bibinfo{person}{Andrew Howes},
  \bibinfo{person}{J\"{o}rg M\"{u}ller}, \bibinfo{person}{Aleksi Ikkala},
  \bibinfo{person}{Miroslav Bachinski}, \bibinfo{person}{Arthur Fleig},
  \bibinfo{person}{Florian Fischer}, {and} \bibinfo{person}{Markus Klar}.}
  \bibinfo{year}{2022}\natexlab{}.
\newblock \showarticletitle{What simulation can do for HCI research}.
\newblock \bibinfo{journal}{\emph{Interactions}} \bibinfo{volume}{29},
  \bibinfo{number}{6} (\bibinfo{date}{Nov.} \bibinfo{year}{2022}),
  \bibinfo{pages}{48–53}.
\newblock
\showISSN{1072-5520}
\href{https://doi.org/10.1145/3564038}{doi:\nolinkurl{10.1145/3564038}}


\bibitem[Ortiz et~al\mbox{.}(2024)]%
        {ortiz_dmc-vb_nodate}
\bibfield{author}{\bibinfo{person}{Joseph Ortiz}, \bibinfo{person}{Antoine
  Dedieu}, \bibinfo{person}{Wolfgang Lehrach}, \bibinfo{person}{J.~Swaroop
  Guntupalli}, \bibinfo{person}{Carter Wendelken}, \bibinfo{person}{Ahmad
  Humayun}, \bibinfo{person}{Guangyao Zhou}, \bibinfo{person}{Sivaramakrishnan
  Swaminathan}, \bibinfo{person}{Miguel L\'{a}zaro-Gredilla}, {and}
  \bibinfo{person}{Kevin Murphy}.} \bibinfo{year}{2024}\natexlab{}.
\newblock \showarticletitle{DMC-VB: a benchmark for representation learning for
  control with visual distractors}. In \bibinfo{booktitle}{\emph{Proceedings of
  the 38th International Conference on Neural Information Processing Systems}}
  (Vancouver, BC, Canada) \emph{(\bibinfo{series}{NIPS '24})}.
  \bibinfo{publisher}{Curran Associates Inc.}, \bibinfo{address}{Red Hook, NY,
  USA}, Article \bibinfo{articleno}{211}, \bibinfo{numpages}{29}~pages.
\newblock
\showISBNx{9798331314385}


\bibitem[Saul et~al\mbox{.}(2015)]%
        {saul2015benchmarking}
\bibfield{author}{\bibinfo{person}{Katherine~R Saul}, \bibinfo{person}{Xiao
  Hu}, \bibinfo{person}{Craig~M Goehler}, \bibinfo{person}{Meghan~E Vidt},
  \bibinfo{person}{Melissa Daly}, \bibinfo{person}{Anca Velisar}, {and}
  \bibinfo{person}{Wendy~M Murray}.} \bibinfo{year}{2015}\natexlab{}.
\newblock \showarticletitle{Benchmarking of dynamic simulation predictions in
  two software platforms using an upper limb musculoskeletal model}.
\newblock \bibinfo{journal}{\emph{Computer methods in biomechanics and
  biomedical engineering}} \bibinfo{volume}{18}, \bibinfo{number}{13}
  (\bibinfo{year}{2015}), \bibinfo{pages}{1445--1458}.
\newblock


\bibitem[Schulman et~al\mbox{.}(2017)]%
        {ppo_schulman}
\bibfield{author}{\bibinfo{person}{John Schulman}, \bibinfo{person}{Filip
  Wolski}, \bibinfo{person}{Prafulla Dhariwal}, \bibinfo{person}{Alec Radford},
  {and} \bibinfo{person}{Oleg Klimov}.} \bibinfo{year}{2017}\natexlab{}.
\newblock \bibinfo{title}{Proximal Policy Optimization Algorithms}.
\newblock
\showeprint[arXiv]{1707.06347}
\urldef\tempurl%
\url{http://arxiv.org/abs/1707.06347}
\showURL{%
\tempurl}


\bibitem[Selder et~al\mbox{.}(2025)]%
        {demystifying2025}
\bibfield{author}{\bibinfo{person}{Hannah Selder}, \bibinfo{person}{Florian
  Fischer}, \bibinfo{person}{Per~Ola Kristensson}, {and}
  \bibinfo{person}{Arthur Fleig}.} \bibinfo{year}{2025}\natexlab{}.
\newblock \showarticletitle{Demystifying Reward Design in Reinforcement
  Learning for Upper Extremity Interaction: Practical Guidelines for
  Biomechanical Simulations in HCI}. In \bibinfo{booktitle}{\emph{Proceedings
  of the 38th Annual ACM Symposium on User Interface Software and Technology}}
  \emph{(\bibinfo{series}{UIST '25})}. \bibinfo{publisher}{Association for
  Computing Machinery}, \bibinfo{address}{New York, NY, USA}, Article
  \bibinfo{articleno}{95}, \bibinfo{numpages}{17}~pages.
\newblock
\showISBNx{9798400720376}
\href{https://doi.org/10.1145/3746059.3747779}{doi:\nolinkurl{10.1145/3746059.3747779}}


\bibitem[Shen et~al\mbox{.}(2015)]%
        {shen_15_effect}
\bibfield{author}{\bibinfo{person}{Zhangfan Shen}, \bibinfo{person}{Chengqi
  Xue}, \bibinfo{person}{Jing Li}, {and} \bibinfo{person}{Xiaozhou Zhou}.}
  \bibinfo{year}{2015}\natexlab{}.
\newblock \showarticletitle{Effect of Icon Density and Color Contrast on Users'
  Visual Perception in Human Computer Interaction}. In
  \bibinfo{booktitle}{\emph{Engineering Psychology and Cognitive Ergonomics}},
  \bibfield{editor}{\bibinfo{person}{Don Harris}} (Ed.).
  \bibinfo{publisher}{Springer International Publishing},
  \bibinfo{address}{Cham}, \bibinfo{pages}{66--76}.
\newblock
\showISBNx{978-3-319-20373-7}


\bibitem[Shi et~al\mbox{.}(2024)]%
        {crtypist_24_shi}
\bibfield{author}{\bibinfo{person}{Danqing Shi}, \bibinfo{person}{Yujun Zhu},
  \bibinfo{person}{Jussi P.~P. Jokinen}, \bibinfo{person}{Aditya Acharya},
  \bibinfo{person}{Aini Putkonen}, \bibinfo{person}{Shumin Zhai}, {and}
  \bibinfo{person}{Antti Oulasvirta}.} \bibinfo{year}{2024}\natexlab{}.
\newblock \showarticletitle{CRTypist: Simulating Touchscreen Typing Behavior
  via Computational Rationality}. In \bibinfo{booktitle}{\emph{Proceedings of
  the 2024 CHI Conference on Human Factors in Computing Systems}} (Honolulu,
  HI, USA) \emph{(\bibinfo{series}{CHI '24})}. \bibinfo{publisher}{Association
  for Computing Machinery}, \bibinfo{address}{New York, NY, USA}, Article
  \bibinfo{articleno}{942}, \bibinfo{numpages}{17}~pages.
\newblock
\showISBNx{9798400703300}
\href{https://doi.org/10.1145/3613904.3642918}{doi:\nolinkurl{10.1145/3613904.3642918}}


\bibitem[Slaoui et~al\mbox{.}(2019)]%
        {slaoui_robust_19}
\bibfield{author}{\bibinfo{person}{Reda~Bahi Slaoui},
  \bibinfo{person}{William~R. Clements}, \bibinfo{person}{Jakob~N. Foerster},
  {and} \bibinfo{person}{S{\'{e}}bastien Toth}.}
  \bibinfo{year}{2019}\natexlab{}.
\newblock \bibinfo{title}{Robust Domain Randomization for Reinforcement
  Learning}.
\newblock
\showeprint[arXiv]{1910.10537}
\urldef\tempurl%
\url{http://arxiv.org/abs/1910.10537}
\showURL{%
\tempurl}


\bibitem[Wang et~al\mbox{.}(2025)]%
        {wang2025deep}
\bibfield{author}{\bibinfo{person}{Shixiao Wang}, \bibinfo{person}{Runsheng
  Zhang}, \bibinfo{person}{Junliang Du}, \bibinfo{person}{Ran Hao}, {and}
  \bibinfo{person}{Jiacheng Hu}.} \bibinfo{year}{2025}\natexlab{}.
\newblock \showarticletitle{A Deep Learning Approach to Interface Color Quality
  Assessment in HCI}. In \bibinfo{booktitle}{\emph{2025 8th International
  Symposium on Big Data and Applied Statistics (ISBDAS)}}.
  \bibinfo{publisher}{Institute of Electrical and Electronics Engineers
  (IEEE)}, \bibinfo{address}{Guangzhou, China}, \bibinfo{pages}{416--420}.
\newblock
\href{https://doi.org/10.1109/ISBDAS64762.2025.11117032}{doi:\nolinkurl{10.1109/ISBDAS64762.2025.11117032}}


\bibitem[Wang et~al\mbox{.}(2021)]%
        {Wang_2021_CVPR}
\bibfield{author}{\bibinfo{person}{Xudong Wang}, \bibinfo{person}{Long Lian},
  {and} \bibinfo{person}{Stella~X. Yu}.} \bibinfo{year}{2021}\natexlab{}.
\newblock \showarticletitle{Unsupervised Visual Attention and Invariance for
  Reinforcement Learning}, In \bibinfo{booktitle}{Proceedings of the IEEE/CVF
  Conference on Computer Vision and Pattern Recognition (CVPR)}.
\newblock \bibinfo{journal}{\emph{2021 IEEE/CVF Conference on Computer Vision
  and Pattern Recognition (CVPR)}}  \bibinfo{volume}{2021},
  \bibinfo{pages}{6673--6683}.
\newblock
\urldef\tempurl%
\url{https://api.semanticscholar.org/CorpusID:233168943}
\showURL{%
\tempurl}


\bibitem[Xiang et~al\mbox{.}(2024)]%
        {simuser_xiang_24}
\bibfield{author}{\bibinfo{person}{Wei Xiang}, \bibinfo{person}{Hanfei Zhu},
  \bibinfo{person}{Suqi Lou}, \bibinfo{person}{Xinli Chen},
  \bibinfo{person}{Zhenghua Pan}, \bibinfo{person}{Yuping Jin},
  \bibinfo{person}{Shi Chen}, {and} \bibinfo{person}{Lingyun Sun}.}
  \bibinfo{year}{2024}\natexlab{}.
\newblock \showarticletitle{SimUser: Generating Usability Feedback by
  Simulating Various Users Interacting with Mobile Applications}. In
  \bibinfo{booktitle}{\emph{Proceedings of the 2024 CHI Conference on Human
  Factors in Computing Systems}} (Honolulu, HI, USA)
  \emph{(\bibinfo{series}{CHI '24})}. \bibinfo{publisher}{Association for
  Computing Machinery}, \bibinfo{address}{New York, NY, USA}, Article
  \bibinfo{articleno}{9}, \bibinfo{numpages}{17}~pages.
\newblock
\showISBNx{9798400703300}
\href{https://doi.org/10.1145/3613904.3642481}{doi:\nolinkurl{10.1145/3613904.3642481}}


\bibitem[Zhu et~al\mbox{.}(2024)]%
        {theory_of_mind_zhu_24}
\bibfield{author}{\bibinfo{person}{Yifan Zhu}, \bibinfo{person}{Hannah
  VanderHoeven}, \bibinfo{person}{Kenneth Lai}, \bibinfo{person}{Mariah
  Bradford}, \bibinfo{person}{Christopher Tam}, \bibinfo{person}{Ibrahim
  Khebour}, \bibinfo{person}{Richard Brutti}, \bibinfo{person}{Nikhil
  Krishnaswamy}, {and} \bibinfo{person}{James Pustejovsky}.}
  \bibinfo{year}{2024}\natexlab{}.
\newblock \showarticletitle{Modeling Theory of Mind in Multimodal HCI}. In
  \bibinfo{booktitle}{\emph{Human-Computer Interaction}},
  \bibfield{editor}{\bibinfo{person}{Masaaki Kurosu} {and}
  \bibinfo{person}{Ayako Hashizume}} (Eds.). \bibinfo{publisher}{Springer
  Nature Switzerland}, \bibinfo{address}{Cham}, \bibinfo{pages}{205--225}.
\newblock
\showISBNx{978-3-031-60405-8}


\end{thebibliography}

%\clearpage
\appendix
%\section{Robustness in the Presence of Distractors}
%The figures below illustrate the robustness results for static and moving distractors.
\twocolumn[{%
\section{Illustrations of Robustness in the Presence of Distractors}
}]
\begin{figure*}[b]
    \centering
    \includegraphics[width=\linewidth]{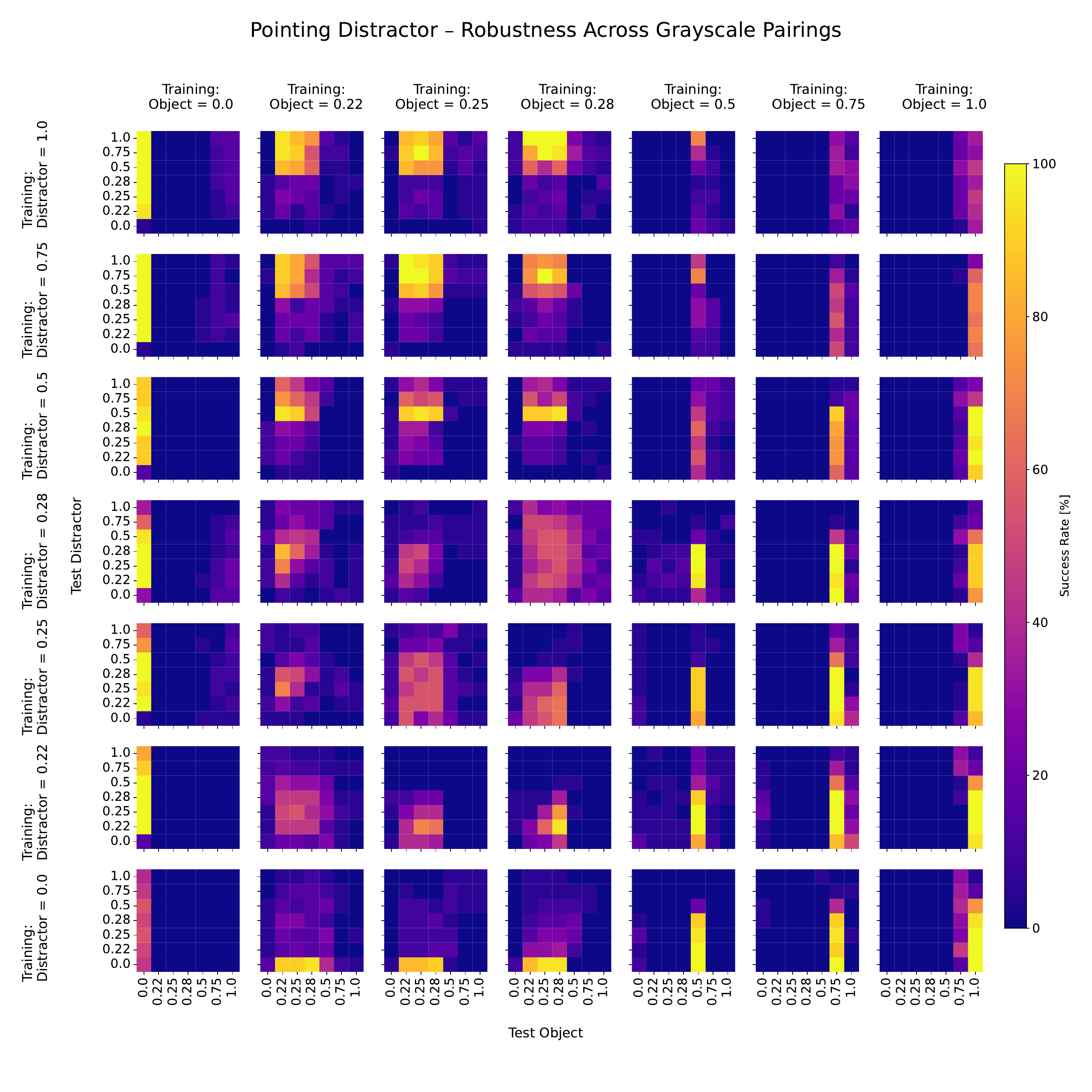}
    \caption{Robustness across static distractor–object luminance combinations for pointing. Agents trained with object luminance similar to the background (0.25) show higher robustness, generalizing well to evaluation conditions with luminances close to those seen during training.}
    \label{fig:pointing_distractor_robustness_heatmap_reduced}
    \Description{The figure shows how agents trained on specific static distractor-object luminance pairings generalize to other luminance combinations for the pointing task. It consist of 49 individual heatmaps representing the performance of one agent trained on a specific static distractor-object luminance combination evaluated across different test luminance combinations. Rows indicate the background luminance, and columns show object luminance. Success rates range from 0 to 100\% and agents trained on settings with object luminances close to the background luminance achieve best robustness.}
\end{figure*}

\begin{figure*}[hb]
    \centering
    \includegraphics[width=\linewidth]{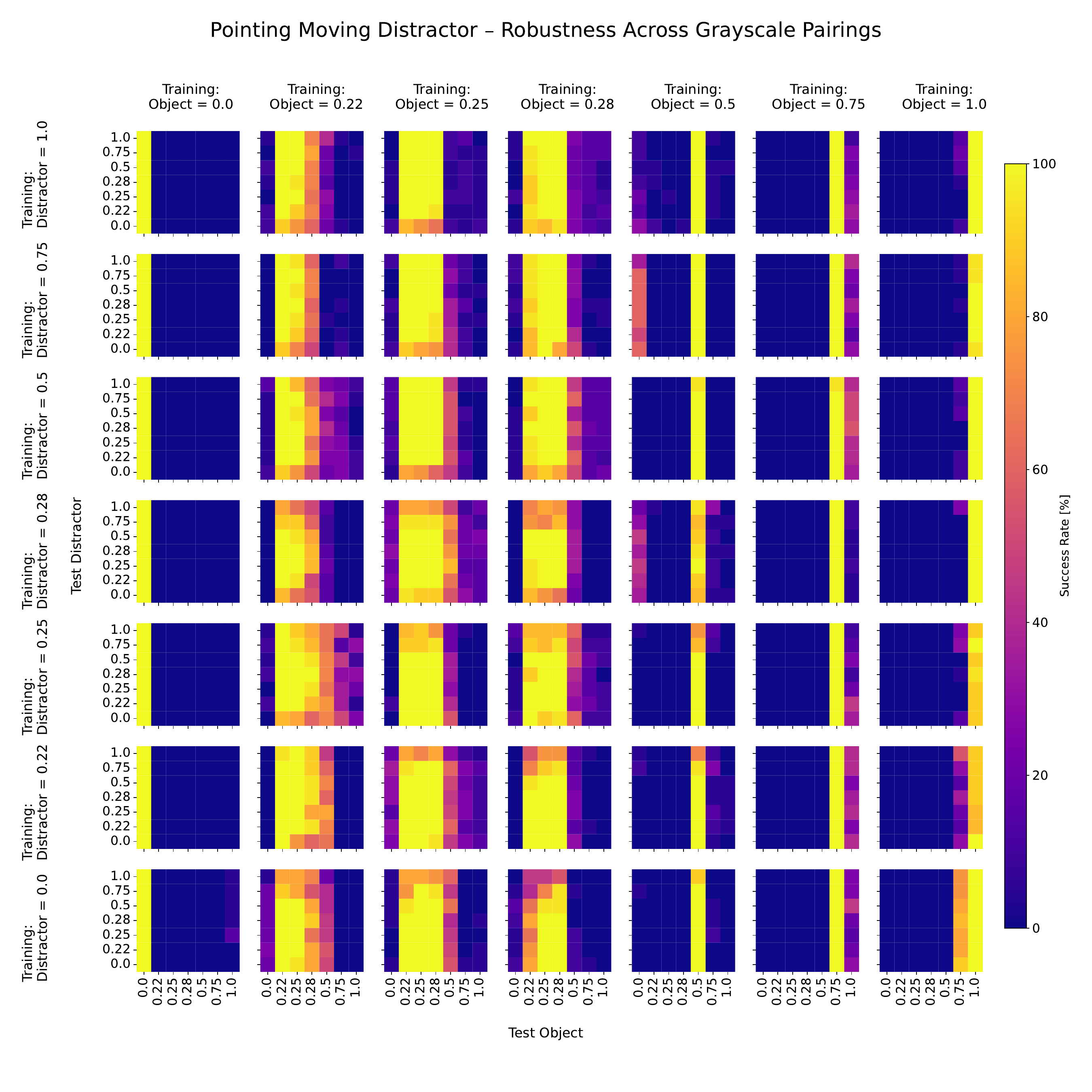}
    \caption{Robustness across moving distractor-object luminance combinations for pointing. Agents trained on objects luminances that are similar to the background luminance (0.25) achieve good robustness when evaluated with luminances similar to training luminances.}\label{fig:pointing_moving_distractor_robustness_heatmap_reduced}
    \Description{The figure shows how agents trained on specific moving distractor-object luminance pairings generalize to other luminance combinations for the pointing task. It consist of 49 individual heatmaps representing the performance of one agent trained on a specific moving distractor-object luminance combination evaluated across different test luminance combinations. Rows indicate the background luminance, and columns show object luminance. Success rates range from 0 to 100\% and agents trained on settings with object luminances close to the background luminance achieve best robustness.}
\end{figure*}

\begin{figure*}
        \centering
        \includegraphics[width=\linewidth]{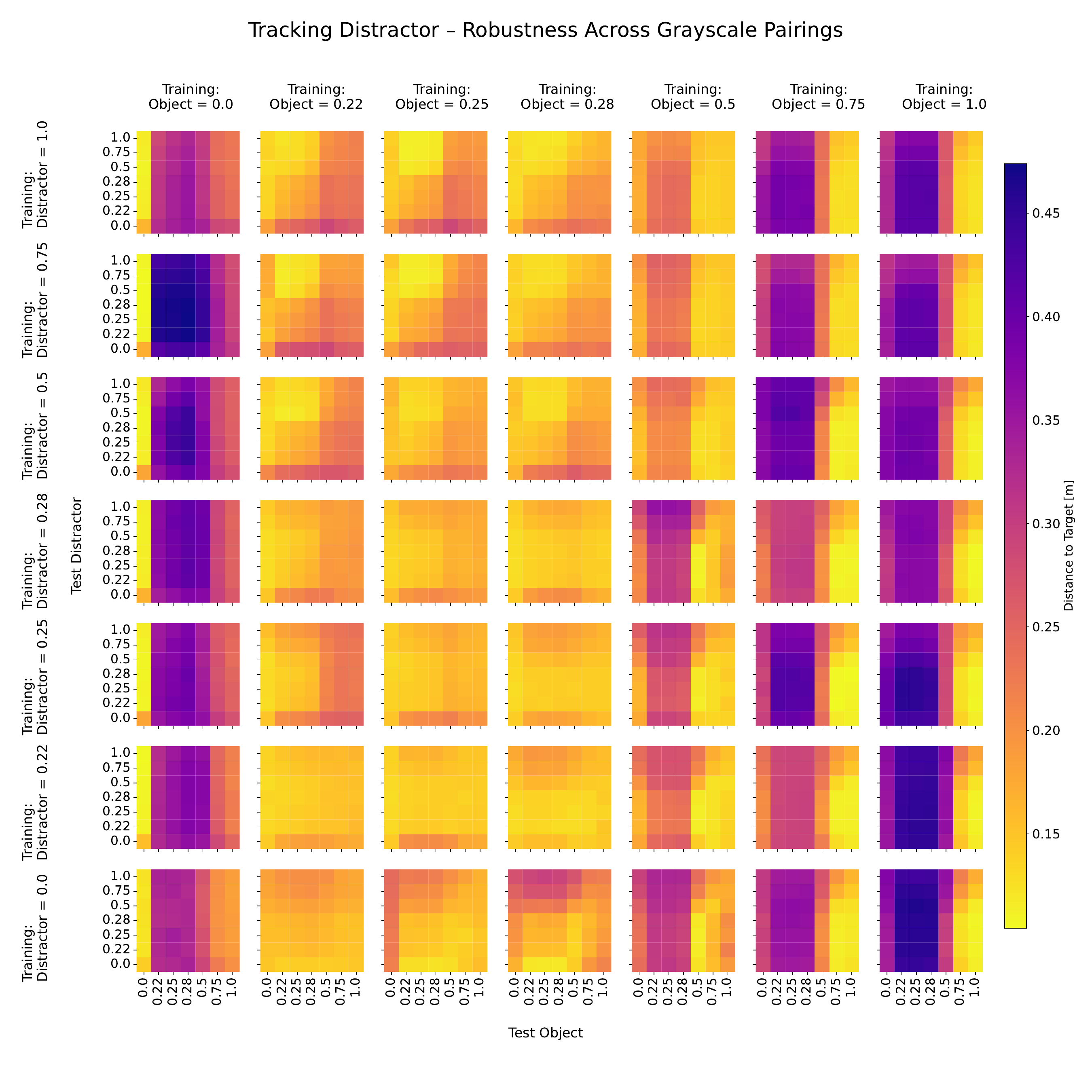}
        \caption{Robustness across static distractor-object luminance combinations for tracking. Agents trained on objects luminances that are similar to the background luminance (0.25) achieve good robustness.}
        \Description{The figure shows how agents trained on specific static distractor-object luminance pairings generalize to other luminance combinations for the tracking task. It consist of 49 individual heatmaps representing the performance of one agent trained on a specific static distractor-object luminance combination evaluated across different test luminance combinations. Rows indicate the background luminance, and columns show object luminance.
        Distances to target range from 0.1 m to 0.47 m and agents trained on settings with object luminances close to the background luminance achieve best robustness.}
        \label{fig:tracking_distractor_robustness_heatmap_reduced}
    \end{figure*}

    \begin{figure*}
        \centering
        \includegraphics[width=\linewidth]{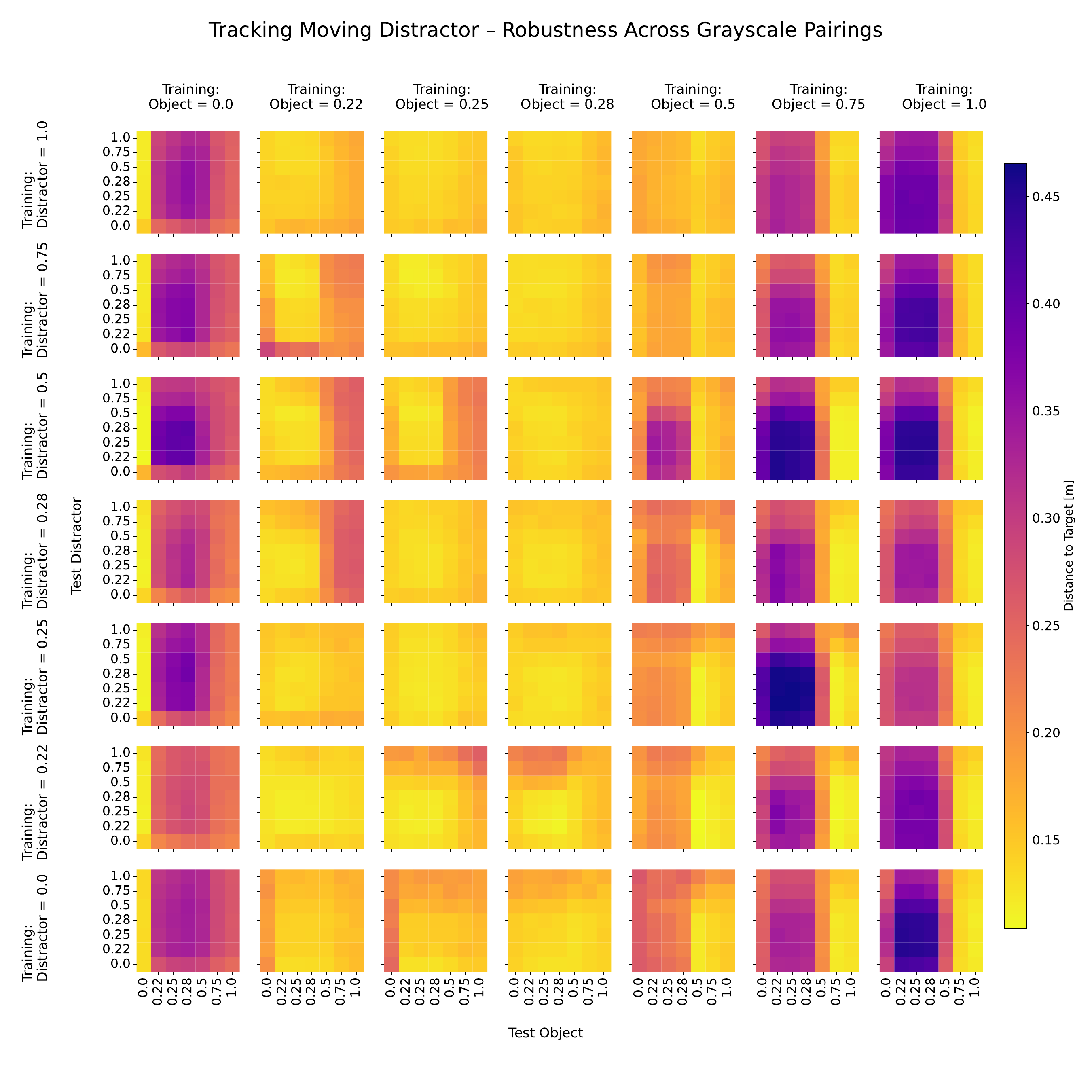}
        \caption{Robustness across moving distractor-object luminance combinations for tracking. Agents trained on objects luminances that are similar to the background luminance (0.25) achieve good robustness.}
        \Description{The figure shows how agents trained on specific moving distractor-object luminance pairings generalize to other luminance combinations for the tracking task. It consist of 49 individual heatmaps representing the performance of one agent trained on a specific moving distractor-object luminance combination evaluated across different test luminance combinations. Rows indicate the background luminance, and columns show object luminance.
        Distances to target range from 0.1 m to 0.47 m and agents trained on settings with object luminances close to the background luminance achieve best robustness.}
        \label{fig:tracking_moving_distractor_robustness_heatmap_reduced}
        \end{figure*}
\end{document}